\documentclass[a4paper]{article}
\usepackage{amssymb}
\usepackage{graphicx}
\def\ld#1{{{\rm ld}\,{#1}}}
\def\tr#1{{{\rm tr}\,{#1}}}

\begin{document}

\begin{center}

{\bf\Large Algebraic mechanics as an accessible toy model
demonstrating entropy generation from reversible 
microscopic dynamics}

\bigskip\bigskip

{\bf Thomas Fischbacher\\}
\smallbreak
{\em University of Southampton
  School of Engineering Sciences\\
  Highfield Campus\\
  University Road, SO17 1BJ Southampton, United Kingdom\\}
{\small {\tt t.fischbacher@soton.ac.uk}}

\end{center}

\begin{abstract}
\noindent
One observes that a considerable level of confusion remains about some
of those aspects of irreversibility, entropy generation and `the arrow
of time' which actually are well understood.  This demands that great
care must be taken in any discussion of irreversibility to use
clear-cut notions and precise language in order to be definite about
which property follows from which assumption. In this work, a novel
toy model of `algebraic mechanics' is presented that elucidates
specific key aspects of entropy generation in a system with extremely
simple reversible fundamental dynamics. It is argued why insights
gained through a detailed quantitative study of this toy model also
have to be taken into account for any realistic model of microscopic
dynamics, classical or quantum alike. As irreversibility also touches
upon the quantum mechanical measurement process (through the `proof'
of the `H-Theorem'), a simple way to address the tenacious question
`when (and how) the wave function collapses' is offered.
\end{abstract}

\section{Introduction}

One of the most fundamental observations is that most processes we
experience daily are intrinsically irreversible (`one cannot make
grain from beer'). On the other hand, the fundamental laws governing
the physics of the building blocks of our world -- most importantly,
gravity and quantum electrodynamics -- feature time reversal
symmetry. So, how can reversible microscopic behaviour give rise to
irreversible collective macrophysical phenomena? When discussing this
question, the situation is actually obscured by the interplay of a
number of very different facets: for one, we know that the standard
model of particle physics contains exotic CP-violating processes which
(due to the CPT-theorem) must also violate time reversal symmetry,
most notably the decay of neutral K-mesons. While these processes must
have played an important role in the early Universe, in particular to
establish the matter/antimatter asymmetry -- if God would switch them
off right at this instant by pressing a button, we would not expect
this to have any consequence whatsoever anymore towards the observed
asymmetry between `forward in time' and `backward in time'. Another
important aspect is that the expanding Universe has a dense, hot past,
reaching back to the Big Bang, while it does not seem to have a
corresponding fate in the future.

Important as it is to consider cosmological and field theoretic facets
of the question of the nature of time asymmetry, this must not obscure
that a number of crucial insights already can be obtained by studying
properties of simple physical toy models: if some general observation
can be shown to already follow from very simplistic assumptions, then
the corresponding mechanisms also have to be taken into account when
discussing irreversibility in the context of more realistic physical
models. Furthermore, misconceptions about fundamental issues can
easily give rise to serious misunderstandings in more involved
situations.

\section{Algebraic Mechanics}

A toy model used to study aspects of irreversibility in systems with
reversible microscopic physics should have a number of highly
desirable characteristics:

\begin{itemize}

\item It should be governed by very simple fundamental processes that 
      are \emph{manifestly} time-reversal-symmetric.

\item It should be possible to easily simulate the evolution of the system
      on a computer at no loss of accuracy due to discretisation
      errors or similar technological restrictions.

\item It should be feasible to simulate a potentially large number of time steps 
      with constant memory requirements.

\item The number of states of the system should be large, but finite, and ideally,
      entropy computations should not involve any mathematics more
      advanced than high school level.

\item The system must provide some model for the identification of macroscopic
      properties from microscopic configurations.

\end{itemize}

One system that nicely satisfies all these conditions is a lattice gas
model of discernible particles in which the fundamental interaction is
scattering between particles. This model furthermore is made exactly
tractable numerically (i.e. free of accumulating rounding errors) by
taking particle positions and velocities not to be real numbers, but
elements of the finite field $\mathbb{F}_p$, with $p$ being a
prime. For all the examples in the rest of this work, we will
specifically choose $p=19$. While this trick simultaneously makes the
number of microstates of the system finite, the resulting model then
unfortunately only retains a formal resemblance to real world
physics. Still, once the important insights have been established
utilizing this model as a tool, one easily convinces oneself that many
relevant properties can be lifted directly e.g. to a model of the
dynamics of hard spheres. Major technical restrictions are that
numerical computations then have to be done with ridiculously high
precision, which also will depend on the length of the interval of
time to be simulated. On the conceptual side, going to hard spheres
will require replacing the simple counting of states
(i.e. combinatorics) with more advanced measure theory.

These three rules provide an axiomatic specification of the behaviour
of the `algebraic mechanics' model:

\begin{itemize}

\item (Arithmetics) All arithmetics is to be done in the field $\mathbb{F}_p$, i.e. 
      modulo the prime $p$.\footnote{Readers not experienced with working
      over finite number fields are reminded that, for $p$ prime, division
      is a well-defined operation, as there will be precisely one $x$ 
      satisfying $a\cdot x = b\;{\rm mod}\, p$ for $0\neq a\in \mathbb{F}_p$.}
      In the following, these arithmetic operations will be denoted by
      $+_p, -_p, \cdot_p$, etc.

\item (World) The system consists of finitely many labeled (i.e. discernible) particles
      living on the cells of a two-dimensional $p\times p$
      board. Their physical degrees of freedom hence consist of two position coordinates
      $(x,y)\in \mathbb{F}_p\times \mathbb{F}_p$ as well as two velocity coordinates
      $(v_x,v_y)\in \mathbb{F}_p\times \mathbb{F}_p$. Multiple particles may occupy the
      same site on the lattice.

\item (Dynamics) Time advances in discrete steps. A single step consists of $p$ subsequent 
      stages, where each stage consists of three subsequent phases: Motion, Scattering, Motion.
      In a `Motion' phase, every particle's coordinates are increased according to the
      particle's velocities: $(x_{(n)}',y_{(n)}')=(x_{(n)}+_pv_{x,(n)},y_{(n)}+_pv_{y,(n)})$.
      In a `Scattering' phase, particles' positions are not updated, but whenever multiple
      particles occupy the same cell, their average velocity $\bar v$ is determined
      (using $\mathbb{F}_p$ arithmetics), and the velocity $v_{(n)}=\bar v+_p\Delta v_{(n)}$ 
      of every particle in that cell is then replaced by $v_{(n)}'=\bar v-_p\Delta v_{(n)}$.
      If $p$ or more particles occupy the same cell, no scattering happens in that cell
      (i.e. velocities are not changed).
\end{itemize}

One immediately notices that:

\begin{itemize}

\item A scattering phase does not change the average velocity of all particles occupying
      the same cell, hence `total momentum is conserved'.

\item A stage is made up in such a way of phases that it (a) is inherently
      time-reversal-symmetric, and (b) involves changes to both positions and velocities.

\item The rule to exclude scattering for cells containing $p$ or more particles manages to
      make the dynamics well-defined in every situation and retains interesting nontrivial
      scattering properties as long as the number of particles is not far larger than
      the number of cells.

\item Systems which have been set up in such a way that scattering events do not take
      place -- e.g. one particle per row, all of them moving horizontally only -- return
      to their initial configuration after $p$ stages, i.e. one time step. Hence,
      under these rules, the time evolution of the system is governed by two- and
      multiparticle scattering processes.

\end{itemize}

The rules given here are simple enough to be easily implemented in
Emacs Lisp, so that everybody's favorite text editor can be used to
study the behaviour of the system. A short piece of program code that
implements a complete simulation framework is shown in
appendix~\ref{elisp}.

When starting from a very specific initial condition, such as a number
of particles arranged as a compact block, with random initial
velocities, one finds that for small block sizes, scattering processes
are so rare that not much interesting happens on reasonable time
scales. For our choice $p=19$, a $9\times 9$ block appears to be just
of the right size to show interesting dynamics, as demonstrated in
figure~\ref{ex:dynamics9x9}.

\begin{figure}
\begin{center}
{\small
\begin{verbatim}

       T= 0                  T= 1                 T= 2
111111111..........  1111..1.1..........  1111.1111...1......
111111111..........  111.11111..........  111.11111........1.
111111111..........  111211111..........  11111111...........
111111111..........  1111.111...........  .111111............
111111111..........  11111111...........  111111111..........
111111111..........  11.111211..........  11.1.11.1..........
111111111..........  1...1.121..........  111.1..1...........
111111111..........  1.211.11...........  11111.11...........
111111111..........  .11111111..........  .1.111111..........
...................  ..............1....  ...................
...................  ....11.............  .....2.............
...................  .1.................  .1...........1....1
...................  ...................  .........1.........
...................  .1...............1.  ...................
...................  ...................  .................1.
...................  1...............1..  1..................
...................  ....1.......1......  ....1........1.....
...................  .............1.....  1............1....1
...................  ..................1  ............1.....1

        T= 5                 T=10                 T=20	
1111.11.1.1........  1.1..1..1.1...1....  ..1..11.........1..
111.21111..........  121.111.1..1.......  ..11...21..1......1
11.111.11..........  1..1...1........1..  ......1.1...2...2..
.1.1.111...........  ..1...1........1.1.  ....2.2..........1.
.1111111...........  ..221111...........  ..1..11....1.......
11.111211..........  111.111.1..........  1....2..1..........
11....12...........  .1..11.............  .1.......11......1.
1.111.11...........  1.111.1.1.........1  1..1.........11....
.11111111..1.......  ..111.1....1....1..  ..11......1......1.
...................  1....1............1  ...........1.2....1
....11...........1.  ...1.1.............  ........1........1.
.1..........1......  .1........1.......1  .1..1.1...1....1...
.............1.....  ......1.1....1.....  11.....1..1..2...1.
.1.................  .1................1  1..................
..............11...  ........1.......11.  .1...............1.
1...............1..  .1...........11....  ...12...1....1.....
....1..............  .......1..21.1..1..  .1.....1..1.11.....
.............1....1  ....1..............  ...1...1....1..1...
.......1..........1  ............1......  1..............3...
\end{verbatim}}

\end{center}
\caption{\label{ex:dynamics9x9} Time evolution of a $9\times 9$ block of particles (example).
Numbers denote the number of particles in a given cell, a dot denotes
an empty cell.}
\end{figure}

In systems in which the {\em ergodic hypothesis} holds (or at least a
weakened version thereof which claims that the system will trace out a
substantial fraction of the accessible configuration space in an
`effectively chaotic' manner), asking the question what microstate the
system is in at individual points in `macroscopic time' (i.e. the
scale of time differences is considerably larger than the time scale
of microscopic processes) is equivalent to obtaining data from an
uncorrelated source of randomness (such as a perfect die). Then, the
Shannon entropy of such a random process just corresponds to the
Boltzmann entropy of the physical system (up to a dimensionful
proportionality constant required to match the statistical
interpretation with the phenomenology of macroscopic thermodynamics).

Unfortunately, in a number of situations, the ergodic hypothesis is
much more attractive than justifiable. In particular, we can consider
ourselves lucky that the solar system does not seem to trace out all
the mechanically possible configurations that would be allowed taking
only the classical conservation laws into account. At first, this
observation seems to be a major hurdle for the construction of a
general theory explaining macroscopic phenomena in terms of
microscopic processes. We will see, however, that the desirable link
between Shannon and Boltzmann entropy can still be maintained even
without invoking the ergodic hypothesis, if one is willing to pay a
price in the form of a modified interpretation of macroscopic entropy.

\subsection{The entropy of a source of randomness}

If one had to define entropy in but a single sentence, then the
statement that \emph{`Entropy is a linearly additive measure of the
size of a space of possibilities'} presumably would be a strong
contender: while being simple enough to be directly applied to a
number of systems that can even be studied at school level -- such as
casting the die\footnote{Ambrose Bierce would point out here that die
are not cast, but cut.} -- it still contains all the relevant essence
necessary to evolve the analytic formula for entropy both in coding
theory and statistical mechanics, applying not much more than simple
consistency considerations as well as quite elementary mathematics. In
particular, Shannon's entropy formula~\cite{Shannon} is easily derived
from consequent application of the three ideas that `rolling two
perfect die produces twice as much randomness in every step as rolling
just one dice', `rolling five perfect die produces slightly less
randomness than throwing thirteen perfect coins', and `an imperfect
dice that only rolls 1 and 2, with 50\% probability each, is as a
source of randomness equivalent to a perfect coin'. In particular, the
entropy associated to some specific outcome is proportional to the
logarithm of its rarity (inverse probability), and has to be weighted
with the probability. Here, the logarithm ensures that entropy is
additive when composing two independent systems, where the space of
possible configurations grows multiplicatively. A useful choice of
normalization is to associate an entropy of 1 to a perfect coin,
denoting this amount of entropy a `bit'. This boils down to using the
\emph{logarithmus dualis} (base-2 logarithm) $\log{}_2=\ld{}$ when defining 
entropy:

\begin{equation}
S= \sum_i p_i\, \ld 1/p_i = \sum_i -p_i\, \ld p_i
\end{equation}

One of the beauties of the `algebraic mechanics' model is that 
we can easily compute the entropy as the logarithm of the number
of microstate configurations that belong to a macrostate.

Considering a collection of $9\times 9=81$ labeled (i.e. discernible)
particles moving on a $19\times 19$ lattice, the most generic
macrostate description, which does not provide any additional
constraints beyond this, can be realized through $\left(p^4\right)^{81}$
different microstates, as every particle can have arbitrary position
and momentum, both being a pair of mod-p integers. The base-2
logarithm of this number gives the entropy of this macrostate, which
is for $p=19$ just $4\cdot81\cdot \ld 19\approx1376.328514\,{\rm bit}$.


It is extremely important to note here that every constraint on the
configuration of the particles can be translated to a set of
microstates satisfying that constraint, so \emph{entropy is a property
of a macroscopic \emph{description} of a system, not the system
itself!} This means that different observers, which speak about the
\emph{same} system (i.e. microstate), but have a different degree of
information about it, will associate different entropy to it. To make
this point explicitly clear, let us consider the simple geometric
pattern underlying the $T=0$ `initial' configuration in
figure~\ref{ex:dynamics9x9}. We will call this configuration $C_0$.
Descriptions of $C_0$ of different level of detail correspond to
different macrostates, hence different associated entropies:

\begin{itemize}

\item (MS1) A `blind' observer who does not know anything about this
      configuration -- except that ``\emph{it contains 81 labeled particles}''
      -- will associate an entropy of 
      $81\cdot4\cdot\ld{19}\approx1376.33\,{\rm bit}$ to it.

\item (MS2) An observer who describes this configuration as ``\emph{81 particles arranged
      in a regular $9\times 9$ pattern in the top left corner of the lattice,
      with unspecified velocities}'', will associate to it an entropy of
      $81\cdot 2 \cdot \ld{19} + \ld{81!} \approx 1089.33\,{\rm bit}$.

\item (MS3) An observer describing the configuration as ``\emph{all 81 particles
      being located somewhere in the top left $9\times 9$ corner of the lattice,
      with unspecified velocities}'' would associate to this description
      an entropy of 
      $81\cdot 2 \cdot \ld{19}+81\cdot 2\cdot \ld{9} \approx 1201.69\,{\rm bit}$.


\item (MS4) An observer possessing detailed knowledge that ``\emph{the first particle goes
      into the top-left corner, the second into column 2 in the first row, etc.,
      but with unspecified velocities}'', would associate to his description of
      the system the entropy $81\cdot 2 \cdot \ld{19}\approx688.16\,{\rm bit}$.

\item (MS5) An observer having ``\emph{detailed knowledge of the position and velocity
      (i.e. `as specified in the example program') of each individual particle }''
      would associate to this description an entropy of $0\,{\rm bit}$.

\item (MS6) An observer using data-reducing measuring devices that probe spatially averaged
      properties, such as in particular cumulative particle numbers in $3\times 3$
      blocks (resp. $1\times 3$, $3\times 1$, $1\times 1$ blocks for the last row
      and column in the $19\times 19$ lattice), would see ``\emph{nine particles in each
      of the $3\times3$ top left blocks, and none in other positions}''. Such a
      description would be associated to an entropy of 
      $81\cdot 2 \cdot \ld{19}+\ld{\left(\left(9^9\right)^9\frac{81!}{9!^9}\right)}\approx1179.87\,{\rm bit}$.


\end{itemize}

It is especially this last case we will from now on be most concerned with.

\subsection{Entropy generation over time}

Starting from a configuration such as the one named
{\small\texttt{config-a0}} in the code example in the appendix,
scattering processes will soon eradicate all visible
structure. However, fundamental laws being explicitly time symmetric,
we can always `respool' the dynamics by just reversing all
velocities. As there is a $1{:}1$~mapping between microstates at $T=0$
and microstates at any other time $T=T_n$ provided by time evolution,
all that happens in this model is that easily visible spatial
correlation is shifted to and mixed into more complicated correlations
which are completely non-obvious to the human eye (not to speak of the
fact that half the relevant information is missing in plots that do
not show velocities). Experimenting with algebraic mechanics, one
finds that -- while we see perfect reversibility of time evolution --
recurrence phenomena that reproduce initial configurations after an
unexpectedly small number of time evolution steps (say, $\le1000$)
nevertheless do not seem to occur (according to computer experiments).

When going from the time $T=T_0$ to any other time $T=T_n$, fully
reversible microscopic dynamics guarantees that the macrostate (MS2)
-- as a set of \hbox{$8.34\cdot10^{327}$} microstates -- evolves into
another set of precisely the same number of microstates. Hence,
entropy does not change with time in this process. This situation is
completely analogous to the situation in classical mechanics, where
Liouville's theorem ensures the conservation of phase space volume. As
there is a very terse textual description of the macrostate (MS2) at
$T=0$, is there a similarly compact `articulated' description of the
macrostate which we get from (MS2) by time evolution? The best we can
do is:

\begin{quote}
(MS2$_{(n)}$) \emph{A configuration which evolved out of a configuration 
that contained 81 particles in a regular $9\times9$ pattern in the top 
left corner of the lattice, with unspecified velocities, by going from 
$T=0$ to $T=T_n$.}
\end{quote}

This linguistic trick demonstrates the actual conceptual idea
underlying the mathematical proofs of entropy conservation in both
classical as well as quantum systems whose dynamics is given by the
Liouville-von Neumann equation. It should be noted that so far, our
reasoning did not depend on whether $T_n$ lies in the future or in the
past of~$T_0$.

The kind of partial information which we have about any specific
macroscopic system depends on the way our measuring devices work. From
this perspective, we will almost exclusively encounter macrostates
such as (MS6) when studying real systems: \emph{The way our measuring
devices work strongly favors some macrostate descriptions over
others}. It is useful to introduce a special notion here: we want to
call a macrostate whose description corresponds to information
obtained(/in principle obtainable) about a set of microstates by
applying some measuring apparatus~$A$ an
\emph{$A$-observed(/observable) macrostate}. This idea allows us to
define an entropy function $S_A$ that maps information obtained with
the apparatus~$A$ (say, a digital camera, or a thermometer, or both
combined, a pair of human eyes, etc.) to the entropy of the macrostate
description of all the microstates compatible with that observation.

Sticking with the example (MS6), the (specific) measuring
apparatus~$A$ that produces data of such form would embed the
$19\times 19$ lattice into a $21\times 21$ lattice by adding two extra
rows and columns to the right and bottom for bookkeeping purposes only
-- i.e. particles are forbidden to go there -- and measure the number
of particles in each of the $7\times 7$ blocks of size $3\times 3$ in
the enlarged lattice. All information about particle identity as well
as velocity is omitted, only locally averaged information about
spatial distribution is determined.\footnote{Strictly speaking,
`neighborhood' is a much weaker concept in algebraic mechanics than
in more realistic models, as there is no notion of `B is farther away
than C from A'. However, there is a concept of `Applying the
translation that maps A to B once again gives C'.} Explicitly, the
number of microstates corresponding to a measurement that gave $N_j$
particles in the $3\times 3$ block $b_j$ is:
\begin{equation}
N=19^{2\cdot 81}\left(\prod_{j=1}^{7\times 7} |b_j|^{N_j}\right)\cdot\frac{81!}{\prod_{k=1}^{7\times 7} N_k!}
\end{equation}
where $|b_j|$ is the number of real cells in the block $b_j$, normally~9,
but for blocks that contain `padding cells' this may be $3$ or $1$.

\begin{figure}
\label{entropy-evolution}
\begin{center}
\includegraphics[height=8cm]{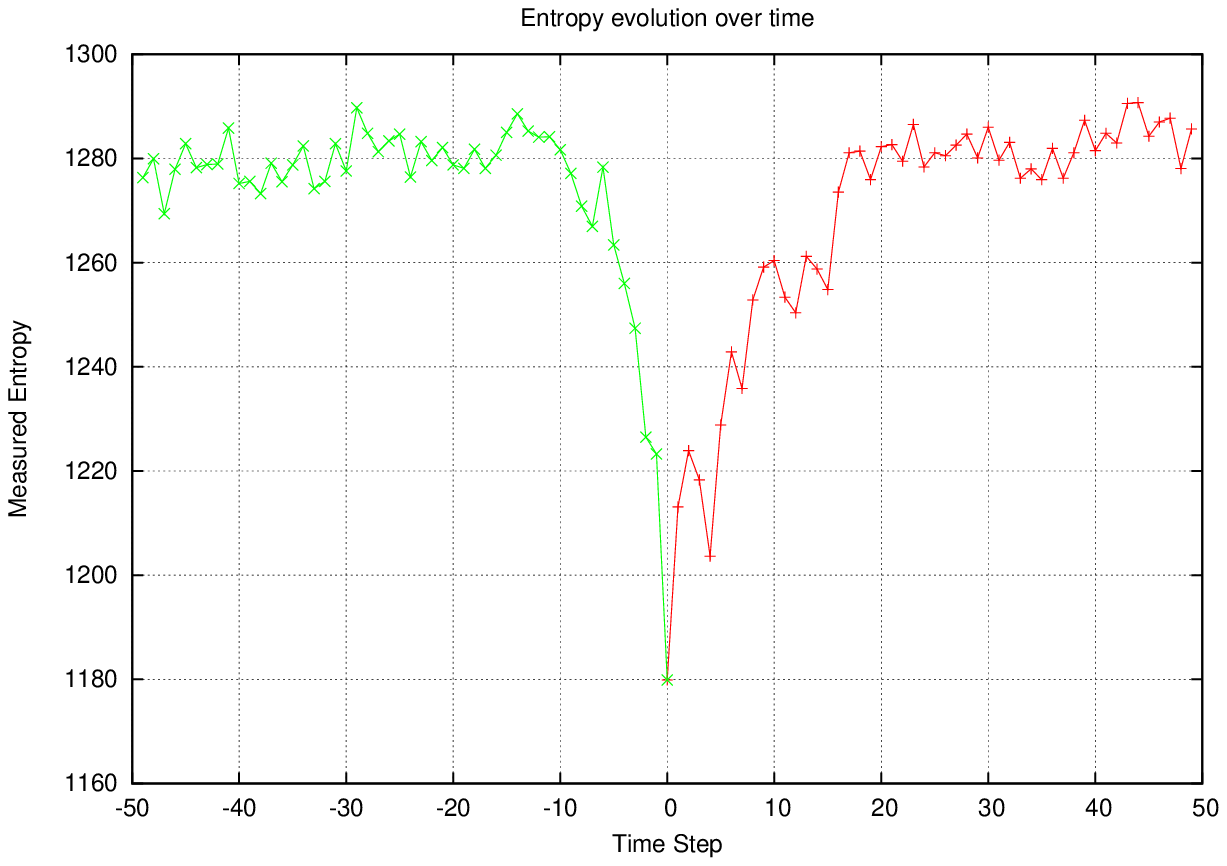}
\end{center}
\begin{caption}{Time evolution of the $A$-observed macrostate entropy.}
\end{caption}
\end{figure}

Time evolution from $T_0$ to $T_n$ maps the configuration (microstate)
$C_0$ to some specific other configuration $C_n$. The evolution of
entropy $S_A(C_n)$ as measured by the apparatus $A$ with time is
displayed in figure~\ref{entropy-evolution}, both for evolution
towards the future as well as evolution towards the past.

Here, it must be pointed out that, while $C_0$ is a very special
configuration amongst all those microstates that belong to the
macrostate (MS6), the behaviour observed is essentially the same for a
generic microstate that belongs to (MS6): entropy increases both
towards the past as well as the future, and also fluctuates. This
claim is easily checked by using the function
\hbox{\small\texttt{random-microstate-compatible-with-macrostate}} from the
code given in the appendix. This produces a random, hence (usually)
generic, microstate corresponding to a given $A$-observed macrostate.

\subsection{Interpretation}

The three crucial ingredients that produce the behaviour shown in
figure~\ref{entropy-evolution} are:

\begin{itemize}
 \item A collection of macrostates that are linked to the behaviour of
 a `data-reducing' measuring device, which reports some kind of reduced
 information about the microstate configuration the system is in.

 \item Microscopic dynamics that is not aligned (i.e. does not
 respect) the data reduction associated with the measuring device.

 \item An initial state which, from the perspective of the measuring
 device, belongs to a macrostate which is special in the sense that
 its observed properties constrain the number of associated microstates
 relative to the number of microstates associated to a generic 
 (again with respect to the measuring device) macrostate.

\end{itemize}

In real systems where the number of degrees of freedom usually is
ridiculously large, the first condition is satisfied automatically by
any conceivable measuring device. Usually, both omission and averaging
over degrees of freedom are involved here. The second condition also
is rather generic. 

So, whenever reversible microscopic dynamics does not care about the
macroscopic notions we use to describe the processes it causes at
macroscopic, `averaged' scales (which practically always is the case),
we expect to encounter the situation that \emph{time evolution does
not change the entropy of the macrostate, but finding a new
description of the evolved macrostate in terms that are discernible by
the measuring device does.} It is this process of `re-articulation' in
which information about the macrostate is lost, and hence entropy
increases. The information lost is precisely that part of the original
knowledge about the system (in this example: averaged positional
information) which -- by the dynamics -- was mixed into more
complicated correlations not detected by the measuring apparatus.

Simply stated, entropy increases whenever we apply dynamics to partial
knowledge about a system and eventually project back onto the same
class of partial system descriptions used initially, unless these
descriptions are compatible with the microscopic dynamics. As there is
no intrinsic reason why they should be, this is normally the case --
but with this criterion in mind, it also is easy to construct
counter-examples (a trivial one being translational motion of a single
solid body). In particular, entropy also increases when we go from the
`present' to the `past': Just as it is difficult to draw conclusions
about the future by studying the present, it is also difficult to draw
conclusions about the past. Generally speaking, if the situation were
so simple that the idea that `entropy increases with time' described
everything there is to know, we should not have much difficulty
answering a question such as `where does the moon come from'? While a
description of a system's past may well correspond to a low entropy
macrostate, we can still easily encounter the situation that, for a
given system, the totality of all conceivable pasts (each of which may
well have low entropy) that are compatible with our knowledge is a
macrostate of higher entropy than the one describing the present
situation.

In this toy system, the magnitude of entropy increase is just a
measure of the amount of information lost by going from a description
such as `evolved out of a system characterized by macroscopic
parameters $P_0$ at time $T_0$' to a description `characterized by
macroscopic parameters $P_n$ at time $T_n$'. Evidently, the latter
macrostate cannot contain fewer microstates than the former: perfect
traceability would mean that we can identify the image of each
microstate under the one-to-one mapping of time
evolution. Less-than-perfect traceability means that the number of
possibilities increases. This is analogous to the very basic
observation that, when adding the numbers 3 and 5, and keeping the
result while forgetting the summands, the number 8 `does not know how
it was produced' -- it may also have been the result of adding 6 to
2. So, in this `irreversible addition', we `produced entropy'.

The observed `fluctuations' on top of the gradual increase in entropy
(which may lead to transient decreases in entropy) correspond to those
situations where the association of macrostates to subsequent
microstates happened to produce a `comparatively small' macrostate
following a larger one. When investigating the dynamics of a single
microstate only, such processes are \emph{a priori} not excluded, and
certainly expected to determine the behaviour of the system at times
far away from $T_0$. If we started instead from an uniformly weighted
collection of all microstates that represent a given macrostate~$M$,
then re-articulation after time evolution would produce a weighted
collection of new $A$-observed macrostates. As a weighted collection of
macrostates again is a macrostate (but usually not an $A$-observable
one), and as this re-articulated description contains extra
microstates in addition to the time-evolution images of the
microstates in~$M$, the entropy of this less stringent description is
larger than the entropy of~$M$: the re-articulation projection loses
information about the time-evolved system. In this sense, the increase
in entropy is inevitable. If we started from any macrostate that is
projected onto a specific $A$-observable macrostate, but contains less
microstates (e.g. only a single one), then there is 
\emph{hidden information} about the system: its state could be known 
more accurately than how it is described by the corresponding
$A$-observable macrostate. The dynamics will mix this hidden
information to a varying degree with those parameters the measuring
device~$A$ is sensitive to, hence giving rise to entropy
fluctuations. Considerations on `Maxwell's Daemon' show that the
magnitude of these fluctuations gives a lower bound on the minimal
effort any physical realization of the measuring device~$A$ must make.

\subsection{Additional remarks}

Putting insights gained by studying the `algebraic mechanics' toy
model into proper physical context requires a few additional remarks.

\subsubsection{Quantum Mechanics, the measurement process, and the `H-Theorem'}

Computing the rate of change of entropy by applying the Liouville-von
Neumann equation to the quantum mechanical expression for entropy
($S=-k_B\tr{\rho\,\ln\rho}$) gives the result that entropy is
conserved. This situation is analogous to the situation encountered in
the toy model when omitting the step of `re-articulation'. One easily
convinces oneself (e.g. by means of an example) that the quantum
mechanical density matrix reduction associated with the measurement
process changes (increases) entropy. This again is associated to an
information-lossy projection, usually called a `quantum jump'. Just as
with the question about entropy generation, there is substantial
confusion about the question `when (and how) the wave function
collapses' (e.g. whether this is a `faster-than-light' process). As
`quantum jumps' lead to an increase of entropy, some relevant aspects
are actually linked, and so are explanations. Considering a combined
system consisting of a `quantum device'~Q (e.g. an excited atom) and a
`measuring device'~D (e.g. a detector), the quantum states of the
combined system are elements of the tensor product Hilbert space
$\mathcal{H}_Q \otimes \mathcal{H}_D$. Due to the interaction between
Q~and~D, the `initial time' quantum states of the form $|q,d\rangle_-$
evolve into entangled linear combinations. In other words, the final
quantum states $|q,d\rangle_+$ we would like to use to describe the
system with \emph{after} the interaction give us a different basis (in
the Heisenberg picture) than that of the initial quantum states. As
one all too often thinks of these quantum states in terms of position
space probability amplitudes, where they look the same, it is
deceptively easy to get confused by failing to discern between
elements of the `initial' Hilbert space $\mathcal{H}_-$ and the
`final' Hilbert space $\mathcal{H}_+$ (which are isomorphic).
Evidently, every description of a measurement process will at some
point have to make the transition from using quantum states in
$\mathcal{H}_-$ to using quantum states in $\mathcal{H}_+$. It is
precisely at this point
\emph{in the description of the process} where this
transition happens that `the wave function collapses', if it is done
in such a way that phase correlation information is lost through
projection. Regardless of whether this step is consciously articulated
or not, it will have to happen somewhere in every description of a
measurement process. Consequently, the confusion of conceptual levels
associated with the question `whether the collapse of the wave
function is a faster-than-light process' is of the same kind as the
confusion demonstrated by the question-answer combination `why the
maid in Shakespeare's poem \emph{A Lover's Complaint} is pale' /
`because it rhymes with \emph{tale}': Here, the question was posed at
the level of content, while the answer was given at the level of
description.\footnote{Another nice example of such a situation occurs
Steve Meretzky's 1986 interactive novel `Leather Goddesses of
Phobos'\cite{Meretzky}, in which a `T-removing machine' turns a rabbit
into a rabbi. It is, of course, not an intrinsic property but pure
misfortune on the side of the rabbit to have ended up, solely by
chance, being named so similarly to a religious teacher in one
particular language.}

This has to be taken into account when using quantum jumps and Fermi's
Golden Rule to justify a Markov model as a basis for a proof of the
quantum mechanical version of Boltzmann's
`H-Theorem'~\cite{Boltzmann}. (Essentially, this then amounts to
proving entropy generation by assuming entropy generating irreversible
fundamental processes.)

\subsubsection{Physical relevance of the toy model}

The key property of the `algebraic mechanics' model is its perfect
computational traceability without losses related to numerical
limitations. While this helps to simplify a number of arguments, the
relevant reasoning can be lifted naturally to more realistic
descriptions of microscopic physics that cannot avoid the problem of
limited numerical precision. This basically means to wrap up most
statements in constructions such as `If one demands numerical
precision X on initial states, then for the given amount of time to
pass, the following holds within numerical precision Y: \dots' where
the precision-related errors can be made small. Apart from
complicating the discussion, this does not introduce qualitatively new
features. Therefore, the value of the toy model lies in helping to
isolate relevant aspects that lead to important insights from
irrelevant ones.

Additionally, while this is not the subject of this work, there are
situations where information about the behaviour of a `continuous'
system can be gained by studying its behaviour in modulo-$p$
arithmetics. Presumably the most famous example of such a situation is
given by the Grothendieck-Katz conjecture~\cite{Katz1982}.

\section{Conclusion}

The `algebraic mechanics' toy model introduced here is both simple and
powerful enough to elucidate some key aspects of the phenomenon of the
thermodynamic arrow of time in terms accessible to a broad audience.
Other toy models that demonstrate entropy generation based on
reversible microscopic dynamics exist, such as the Kac ring
model~\cite{Kac1959} (also see~\cite{Bricmont1996}). The conceptual
advantage of the very simple `algebraic mechanics' toy model is that
model unifies exact computational traceability with formal similarity
to mechanics. It demonstrates entropy generation with respect to time
evolution towards the future \emph{as well as} towards the past, and
gives interpretations of this phenomenon through the notion of
`re-articulation'.

Any `measurement' involves data reduction. In fact, considering the
situation in quantum computing, it may be appropriate to \emph{define}
`a measurement' as `a data-reducing projection'. If the microscopic
dynamics does not respect (i.e. is agnostic about) the eigenspaces of
this projection, then time evolution followed by re-articulation will
lead to an inevitable loss of information contained in the system's
description. Hence, as soon as we use dynamics to extract useful
information about a system at any other time than the present, we see
an increase in entropy, regardless of whether time evolution was
performed towards the future or towards the past. When asking
questions about the future, we will hence observe entropy to increase
towards the future. When doing forensic analysis, we observe the
opposite phenomenon: Extracting information about how precisely an
accident happened gets increasingly difficult the more time has passed
since.

When discussing entropy generation (and also the quantum mechanical
measurement process), great care has to be taken to discern between
concepts that refer to two different levels: the `fundamental dynamics
level' and the `description level'. While conclusions that can be
obtained by studying this toy model leave open many important
questions about the physical `arrow of time' (e.g. why the universe
has a hot past), it is important to first understand what aspects of
entropy generation already follow from very basic general features --
before more advanced physics can be discussed. 

Strictly speaking, this work says nothing new about physical processes
(e.g. fundamental dynamics). Furthermore, it does not offer new
descriptions of physical processes (i.e. thermodynamics). It does,
however, address some occasionally discussed issues concerning the
descriptions of descriptions of physical processes.

\paragraph{Philosophical aspects}

One important source of confusion in the discussion of the `arrow of
time' is the philosophical question of determinism: is the future
determined by the past? Stated differently, if all information about
the world were contained in a spatial slice, and `dynamics' were
nothing else but some invented funny mathematics on top of such an
initial configuration that meaninglessly maps it to other
configurations, why should e.g. some form of `dynamical laws' be more
`real' than another possible imagined choice? Physically speaking,
asking `whether the world is deterministic or not'
\emph{a priori} is as much a non-question as is the question 
`why anything exists at all': As it is impossible to perform
experiments on `the whole world', the toolbox of physics cannot give
an answer -- the only way to come up with an answer is to first find
out how the question one had in mind should have been phrased
accurately. Evidently, we can ask whether some specific process in a
system we can isolate and experiment with is `deterministic'. If a key
property of all experiments is the separation of the system
measurements are performed on from `something else'\footnote{It is
clear what is meant here in a tabletop experiment. When considering
astronomical experiments, however, one realizes that one cannot just
say `the outer environment'. Still, separation occurs, as our
activities in measuring e.g. the microwave background do not reshape it.},
then the question whether `the world is deterministic' is nonphysical
in precisely the same way as the question `what happens if an
unstoppable force meets an unmoveable object', due to a fundamental
contradiction in the assumptions. Certainly, as abolishing relevant
prejudice is an important prerequisite for gaining insight by means of
the scientific method, discussing the `arrow of time' mandates
overcoming all prejudice on determinism first. While this work
demonstrates that the phenomenon of `observed entropy generation'
already happens in an extremely simple completely deterministic toy
model, and reasons that the underlying mechanisms are basic enough to
generalize to virtually all more realistical physical models, this
does not at all touch the question whether some particular fundamental
physics actually follows deterministic laws or not.

\paragraph{Acknowledgments} The original incentive for this work
came from a request to explain in more detail the physical aspects of
a popular talk on the `thermodynamic arrow of time' given by the
author at the interdisciplinary MinD Congress on `Time' in Nuremberg
(N\"urnberg), Germany, 02.10. -- 05.10.2008. The `algebraic mechanics'
model was developed subsequently in an effort to construct a maximally
simple easily traceable quantitative explanation of a key phenomenon
underlying entropy generation. It hence is a pleasure to thank Martin
Dresler for asking the right question -- namely for asking for a
simple quantitative explanation of entropy generation also accessible
to non-physicists.

\newpage
\appendix
\renewcommand{\theequation}{\Alph{section}.\arabic{equation}}
\renewcommand{\thesection}{Appendix \Alph{section}:}

\section{\label{elisp}Appendix: Emacs Lisp Code}

This piece of code, when loaded into the (X)Emacs editor with:\\
\verb$(byte-compile-and-load-file "amech.el")$\\
allows the simulation of the `algebraic mechanics' toy system:

{\small
\begin{verbatim}
(require 'cl)

(defun v-init (n f)
  (let ((v (make-vector n nil)))
    (dotimes (j n) (setf (aref v j) (funcall f j)))
    v))

(defmacro xlambda (args &rest body)
  `(lambda ,args
     (lexical-let ,(mapcar (lambda (n) (list n n)) args) . ,body)))

(defconst amech-prime 19)

(setf amech-div-table
  (v-init amech-prime
          (xlambda (n)
                   (let ((vn (v-init amech-prime
                                     (xlambda (x) (mod (* x n) amech-prime)))))
                     (position 1 vn)))))

(defun a+ (x y) (mod (+ x y) amech-prime))
(defun a- (x y) (mod (- x y) amech-prime))
(defun a* (x y) (mod (* x y) amech-prime))
(defun a/ (x y) (mod (* x (aref amech-div-table y)) amech-prime))
(defun va+ (vx vy) (cons (a+ (car vx) (car vy)) (a+ (cdr vx) (cdr vy))))
(defun va- (vx vy) (cons (a- (car vx) (car vy)) (a- (cdr vx) (cdr vy))))
(defun va/ (vx n)  (cons (a/ (car vx) n) (a/ (cdr vx) n)))

(defun advance-time (config &optional nr-steps)
  (dotimes (step (or nr-steps 1) config)
    (let* ((v00 '(0 . 0))
           (advance (lambda (p) (cons (va+ (car p) (cdr p)) (cdr p))))
           (new-config-1
            (let* ((c (mapcar advance config))
                   (ht-by-pos (make-hash-table :test 'equal)))
              (dolist (np c)
                (push np (gethash (car np) ht-by-pos nil)))
              (maphash
               (lambda (pos particles)
                 (when (< (length particles) amech-prime)
                   (let* ((v-avg (va/ (reduce #'va+ particles
                                              :initial-value v00 :key #'cdr)
                                      (length particles)
                                      )))
                     (dolist (p particles)
                       (setf (cdr p) (va- v-avg (va- (cdr p) v-avg)))))))
               ht-by-pos)
              c))
           (new-config-2 (mapcar advance new-config-1)))
      (setf config new-config-2))))

(defun reverse-velocities (config)
  (mapcar (lambda (p) (cons (car p) (va- '(0 . 0) (cdr p)))) config))

(defun display-config (config)
  (macrolet ((ic (spec)
                 `(let (($x ,spec))
                    (setf ,spec (if (eql $x ?.) ?1
                                  (int-char (+ 1 (char-int $x))))))))
    (let ((board (v-init amech-prime
                         (lambda (n) (make-string amech-prime ?.)))))
      (dolist (particle config)
        (ic (aref (aref board (cdar particle)) (caar particle))))
      (dotimes (j amech-prime)
        (insert (format "\n%s" (aref board j))))))
  (insert "\n") t)

(defun log2-fakt (n)
  (if (= n 0) 0.0
    (+ (/ (log n) (log 2.0)) (log2-fakt (- n 1)))))

(defun config-entropy (config)
  ;; "Measuring" entropy with the device model described in the main text
  (let* ((nr-blocks (ceiling (/ amech-prime 3.0)))
         (boundary (mod amech-prime 3)) ; never =0!
         (b-sizes
          (let ((v (make-vector (* nr-blocks nr-blocks) 9)))
            (dotimes (j nr-blocks)
              (setf (aref v (+ (* nr-blocks j) nr-blocks -1))
                    (* boundary
                       (/ (aref v (+ (* nr-blocks j) nr-blocks -1)) 3))))
            v))
         (counts (make-vector 49 0)))
    (dolist (p config)
      (let* ((xpos (floor (caar p) 3))
             (ypos (floor (cdar p) 3))
             (nr-cell (+ (* ypos 7) xpos)))
        (incf (aref counts nr-cell))))
    (labels
        ((entropy (nr-cell sum)
           (if (= nr-cell 49)
               (reduce (lambda (sf x) (- sf (log2-fakt x)))
                       counts :initial-value (+ sum (log2-fakt 81)))
             (entropy (+ 1 nr-cell)
                      (+ sum (* (aref counts nr-cell)
                                (/ (log (aref b-sizes nr-cell))
                                   (log 2.0))))))))
      (+ (entropy 0 0.0)
         (* 81 2 (/ (log 19) (log 2)))))))

(defun random-microstate-compatible-with-macrostate (n-per-3x3)
  (let ((config nil) (size (length n-per-3x3)))
    (dotimes (j size)
      (dotimes (k size)
        (let ((n (aref (aref n-per-3x3 j) k)))
          (dotimes (m n)
            (push 
             `((,(+ (* 3 j) (random (min 3 (- amech-prime (* 3 j))))) .
                ,(+ (* 3 k) (random (min 3 (- amech-prime (* 3 k))))))
               . (,(random amech-prime) . ,(random amech-prime)))
             config)))))
    config))

(defconst config-a0
  (if nil
      ;; roll the die to produce a random configuration:
      (let ((k 9))
        (coerce (v-init (* k k)
                        (lambda (n) `((,(floor (/ n k)) .
                                       ,(mod n k)) . (,(random amech-prime) .
                                                      ,(random amech-prime)))))
                'list))
    ;; Use a definite initial configuration:
    '(((0 . 0) 14 . 15) ((0 . 1) 0 . 10) ((0 . 2) 6 . 8) ((0 . 3) 0 . 7)
      ((0 . 4) 10 . 18) ((0 . 5) 16 . 4) ((0 . 6) 13 . 7) ((0 . 7) 12 . 13)
      ((0 . 8) 17 . 1) ((1 . 0) 8 . 9) ((1 . 1) 9 . 8) ((1 . 2) 4 . 17)
      ((1 . 3) 1 . 5) ((1 . 4) 14 . 9) ((1 . 5) 16 . 15)  ((1 . 6) 9 . 12)
      ((1 . 7) 11 . 10) ((1 . 8) 3 . 17) ((2 . 0) 18 . 3) ((2 . 1) 6 . 10)
      ((2 . 2) 2 . 4) ((2 . 3) 15 . 16) ((2 . 4) 11 . 8) ((2 . 5) 10 . 10)
      ((2 . 6) 10 . 18) ((2 . 7) 1 . 0) ((2 . 8) 8 . 6) ((3 . 0) 18 . 9)
      ((3 . 1) 11 . 13) ((3 . 2) 18 . 9) ((3 . 3) 10 . 9) ((3 . 4) 2 . 2)
      ((3 . 5) 0 . 5) ((3 . 6) 0 . 0) ((3 . 7) 9 . 7) ((3 . 8) 10 . 11)
      ((4 . 0) 11 . 6) ((4 . 1) 9 . 4) ((4 . 2) 15 . 1) ((4 . 3) 14 . 6)
      ((4 . 4) 0 . 15) ((4 . 5) 6 . 9) ((4 . 6) 2 . 6) ((4 . 7) 18 . 14)
      ((4 . 8) 0 . 18) ((5 . 0) 4 . 10) ((5 . 1) 9 . 6) ((5 . 2) 13 . 10)
      ((5 . 3) 11 . 14) ((5 . 4) 10 . 1) ((5 . 5) 2 . 2) ((5 . 6) 13 . 14)
      ((5 . 7) 8 . 4) ((5 . 8) 18 . 5) ((6 . 0) 5 . 13) ((6 . 1) 11 . 5)
      ((6 . 2) 10 . 17) ((6 . 3) 15 . 13) ((6 . 4) 5 . 15) ((6 . 5) 8 . 6)
      ((6 . 6) 14 . 12) ((6 . 7) 17 . 5) ((6 . 8) 0 . 11) ((7 . 0) 15 . 12)
      ((7 . 1) 6 . 7) ((7 . 2) 14 . 9) ((7 . 3) 9 . 8) ((7 . 4) 4 . 18)
      ((7 . 5) 12 . 4) ((7 . 6) 4 . 17) ((7 . 7) 17 . 15) ((7 . 8) 4 . 9)
      ((8 . 0) 14 . 0) ((8 . 1) 3 . 0) ((8 . 2) 16 . 11) ((8 . 3) 7 . 11)
      ((8 . 4) 5 . 5) ((8 . 5) 17 . 6) ((8 . 6) 16 . 13) ((8 . 7) 18 . 4)
      ((8 . 8) 2 . 13))))

(defconst config-a1
   ;; use this to check the claim of the genericity of entropy evolution
   ;; over time
  (if nil
    (random-microstate-compatible-with-macrostate
     [[9 9 9 0 0 0 0] [9 9 9 0 0 0 0] [9 9 9 0 0 0 0]
      [0 0 0 0 0 0 0] [0 0 0 0 0 0 0] [0 0 0 0 0 0 0]
      [0 0 0 0 0 0 0]])
  '(((8 . 6) 11 . 1) ((6 . 6) 8 . 12) ((6 . 6) 17 . 0) ((6 . 6) 11 . 3)
    ((8 . 8) 3 . 5) ((8 . 6) 16 . 11) ((8 . 8) 9 . 5) ((6 . 6) 16 . 6)
    ((7 . 8) 12 . 18) ((8 . 5) 3 . 8) ((6 . 5) 11 . 14) ((7 . 3) 0 . 6)
    ((8 . 4) 1 . 10) ((7 . 3) 12 . 11) ((6 . 4) 16 . 2) ((7 . 4) 12 . 17)
    ((7 . 4) 11 . 3) ((8 . 5) 6 . 7) ((8 . 2) 17 . 4) ((8 . 2) 3 . 8)
    ((6 . 1) 18 . 8) ((8 . 2) 6 . 10) ((7 . 2) 18 . 18) ((7 . 2) 14 . 8)
    ((7 . 1) 4 . 4) ((7 . 2) 1 . 1) ((7 . 2) 11 . 8) ((3 . 7) 0 . 13)
    ((4 . 7) 2 . 7) ((4 . 6) 15 . 7) ((5 . 7) 2 . 9) ((4 . 6) 13 . 9)
    ((4 . 7) 10 . 2) ((5 . 7) 7 . 11) ((4 . 6) 15 . 16) ((5 . 8) 4 . 11)
    ((5 . 4) 12 . 14) ((4 . 5) 17 . 6) ((5 . 5) 15 . 11) ((4 . 5) 9 . 10)
    ((5 . 5) 4 . 17) ((4 . 5) 16 . 10) ((5 . 5) 17 . 3) ((5 . 3) 2 . 9)
    ((5 . 5) 12 . 5) ((4 . 2) 3 . 6) ((3 . 0) 5 . 4) ((3 . 1) 13 . 10)
    ((3 . 0) 1 . 13) ((4 . 2) 15 . 12) ((4 . 1) 9 . 7) ((4 . 0) 10 . 16)
    ((5 . 2) 13 . 5) ((3 . 1) 10 . 17) ((1 . 7) 5 . 14) ((1 . 6) 10 . 0)
    ((0 . 7) 4 . 7) ((2 . 8) 3 . 7) ((0 . 7) 2 . 12) ((1 . 8) 5 . 4)
    ((1 . 6) 4 . 14) ((1 . 8) 6 . 15) ((1 . 6) 15 . 5) ((0 . 3) 17 . 3)
    ((1 . 4) 4 . 15) ((1 . 3) 11 . 9) ((1 . 4) 5 . 16) ((0 . 4) 10 . 10)
    ((1 . 5) 13 . 8) ((2 . 3) 0 . 1) ((0 . 4) 5 . 12) ((0 . 4) 16 . 1)
    ((0 . 1) 18 . 13) ((0 . 0) 16 . 13) ((1 . 0) 14 . 14) ((2 . 1) 17 . 2)
    ((0 . 0) 13 . 12) ((1 . 1) 15 . 7) ((2 . 1) 9 . 3) ((1 . 2) 9 . 2)
    ((1 . 2) 11 . 5))))


(defun show-evolution (config &optional n tag offset)
  (dotimes (j (or n 10) t)
    (insert (format "\n====== %3s %3d %8.3f ======"
                    (or tag "T") j (config-entropy config)))
    (display-config config)
    (setf config (advance-time config amech-prime)))
  config)

(defun amech-demo (&optional config n)
  (let ((start-config (or config config-a0))
        (nsteps (or n 50))
        (config-at nil) (config-rat nil) (config-a00 nil))
    (setf config-at (show-evolution start-config nsteps "T+"))
    (setf config-rat (reverse-velocities config-at))
    (insert (format "\n[*****************]"))
    (display-config config-rat)
    (insert (format "[*****************]\n"))
    (setf config-a00 (show-evolution config-rat (+ 1 nsteps) "T-" 1))
    (equal config-a00 start-config)))

(amech-demo)
\end{verbatim}
}

\end{document}